\documentclass[msom,nonblindrev]{informs3} 
\usepackage[ruled, linesnumbered, noend, noline]{algorithm2e}
\usepackage{fix-cm}
\usepackage{graphicx}
\usepackage{mathtools}
\usepackage{lmodern}

\RestyleAlgo{ruled}

\usepackage{soul}
\usepackage{xcolor}

\OneAndAHalfSpacedXII 

\usepackage{natbib}
 \bibpunct[, ]{(}{)}{,}{a}{}{,}%
 \def\bibfont{\small}%

\newcommand{\pr}[1]{\mathbb{P}\left(#1\right)}
\newcommand{\ex}[1]{\mathbb{E}\left[#1\right]}
\SetKwComment{Comment}{/* }{ */}
\SetKwInOut{Input}{Input}\SetKwInOut{Output}{Output}

\newcolumntype{M}[1]{>{\centering\arraybackslash}m{#1}}

\TheoremsNumberedThrough     

\EquationsNumberedThrough    

\MANUSCRIPTNO{MSOM-2023-0442.R3} 

\begin{document}


\RUNAUTHOR{Authors}

\RUNTITLE{Multi-output Extreme Spatial Model for Complex Production Systems}

\TITLE{Multi-output Extreme Spatial Model\\ for Complex Aircraft Production Systems}

\ARTICLEAUTHORS{%
\AUTHOR{Cheolhei Lee}
\AFF{Grado Department of Industrial and Systems Engineering, Virginia Polytechnic Institute and State University,
Blacksburg, Virginia 24061, \EMAIL{cheolheil@vt.edu}}
\AUTHOR{Xing Wang}
\AFF{Department of Mathematics, Illinois State University, Normal, Illinois 61790, \EMAIL{xwang70@ilstu.edu}}
\AUTHOR{Xiaowei Yue}
\AFF{Department of Industrial Engineering, Tsinghua University,
Beijing, 100084 China, \EMAIL{yuex@tsinghua.edu.cn}, 
(Corresponding Author)}
\AUTHOR{Jianguo Wu}
\AFF{Department of Industrial Engineering and Management, Peking University, Beijing 100080 China, \EMAIL{j.wu@pku.edu.cn}}

} 

\ABSTRACT{
\textbf{Problem definition:} Data-driven models in machine learning have enabled efficient management of production systems. However, a majority of machine learning models are devoted to modeling the mean response or average pattern, which is inappropriate for studying abnormal extreme events that are often of primary interest in aircraft manufacturing. Since extreme events from heavy-tailed distributions give rise to prohibitive expenditures in system management, sophisticated extreme models are urgently needed to analyze complex extreme risks. Engineering applications of extreme models usually focus on individual extreme events, which is insufficient for complex systems with correlations. 
\textbf{Methodology/results:}  We introduce an extreme spatial model for multi-output response control systems that efficiently captures the dynamics using a bilinear function on two spatial domains for control variables and measurement locations. Marginal parameter modeling and extremal dependence have been investigated. In addition, an efficient graph-assisted composite likelihood estimation and corresponding computational algorithms are developed to cope with high-dimensional outputs. The application to composite aircraft production shows that the proposed model enables comprehensive analyses with superior predictive performance on extreme events compared to canonical methods. \textbf{Managerial implications:} Our method shows how to use an extreme spatial model for predicting extreme events and managing extreme risks in complex production systems such as aircraft. This can help achieve better quality management and operation safety in aircraft production systems and beyond.}



\maketitle

%

\section{Introduction}\label{se:mesm_intro}
Production systems in modern industries encounter demands for unprecedented precision and adaptiveness in processing due to the increased complexity of products and diversified customer preferences. It leads to increments in control and measurement fidelities for quality assurance. Owing to intrinsic uncertainties in production systems, statistical approaches such as statistical process control (SPC) \citep{montgomery2020introduction} and Stream-of-variation (SoV) \citep{shi2006stream} have been extensively applied to production system management and facilitated remarkable progress in various applications (e.g., tolerance synthesis, system design, sensor allocations, quality management). However, a majority of statistical models in production systems focus on central tendencies (e.g., mean, median, and average pattern) and resort to Gaussian distributions, which may not be realistic for extreme events in practice. Moreover, many defects and anomalies of interest arise from tail distributions as extreme events with characteristics different from the Gaussian distribution, such as skewness and heavy tails.

\begin{figure}
    \FIGURE{
    \includegraphics[width=0.9\columnwidth]{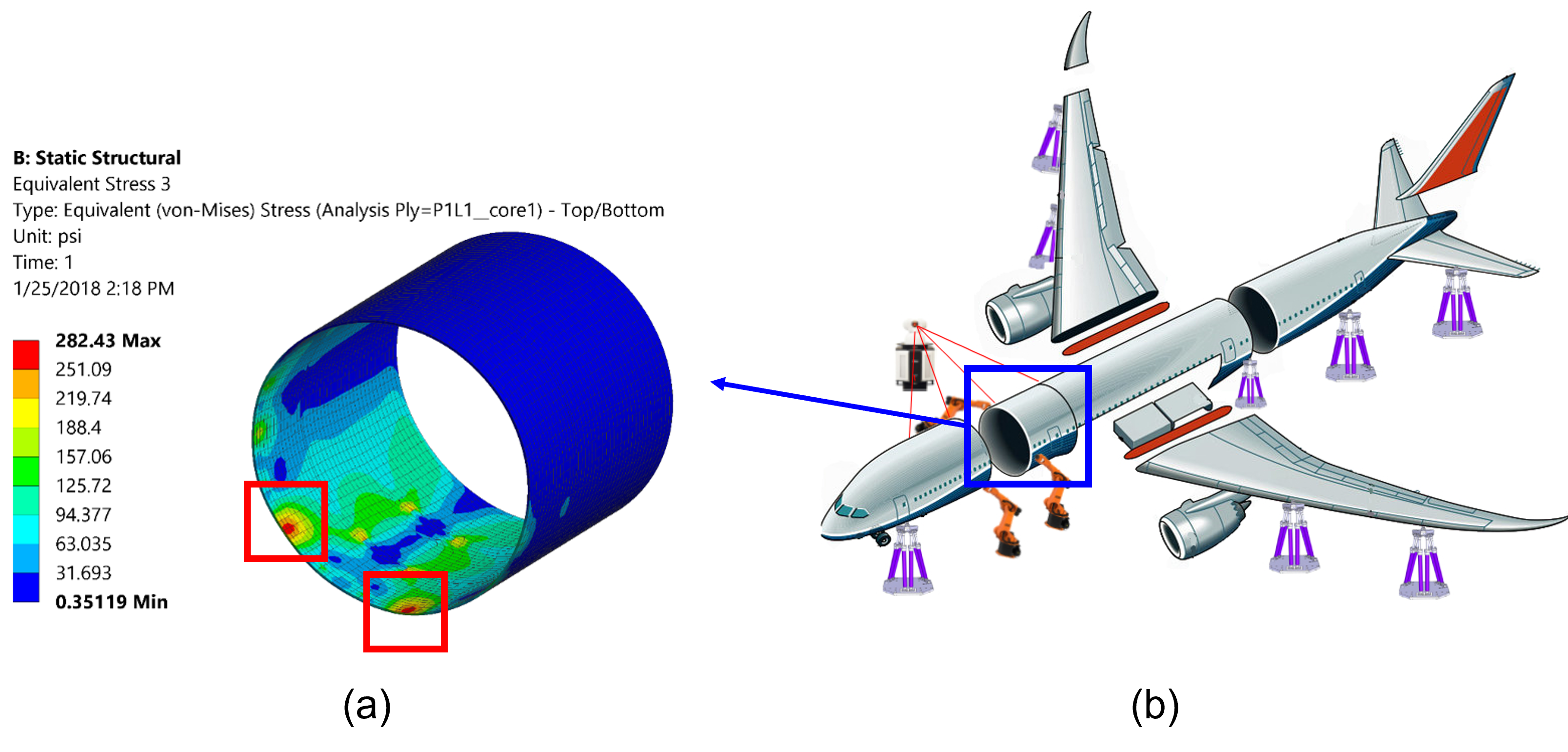}}
    {Extreme risks in aircraft production: (a) potential stress risks marked in red square; (b) composite fuselage assembly in aircraft production \label{fig:aircraft_risk}}{}    
\end{figure}

Managing extreme risks is critical in contemporary production systems since they are rare but cause prohibitive expenditure for system recovery, calibration, or reconstruction. For example, composite fuselage assembly in aircraft production involves shape adjustment because of inevitable dimensional deviations in fuselage sections, while the shape control process with the data-driven deformation model is subject to undesirable structural failures from exceptionally large stress and various uncertainties \citep{yue2018surrogate, lee2022neural}, as shown in Fig. \ref{fig:aircraft_risk}. In additive manufacturing, the quality of a product is significantly influenced by the existence of a large pore that is rarely generated in up-to-date additive manufacturing processes \citep{boyce2017extreme}.

Extreme value analysis has been of great interest to system management and decision analysis \citep{dahan2001extreme,schmidt2022operational,devalkar2018dynamic}, and can be represented with several statistical characteristics. First, statistical inference in extreme value analysis is subject to data scarcity since only a small subset of observations are related to extreme values. Second, extreme value models should be able to forecast further than observed values. It implies that extreme value modeling is not interpolation but extrapolation. Third, modeling multivariate extreme values is challenging since the majority of the density functions have no closed form and require numerical methods to solve. 
 All these distinctions of extreme values and their significance in the real world have led to the development of a focused theory, called extreme value theory (EVT) \citep{coles2001introduction, de2007extreme}.

Conventionally, quantile regression models, referring to specific quantiles, are used to address extreme events in production systems. The quantile models are easy to apply and perform well when the purpose of modeling is to study high- (or low-) quantile values. However, extreme value models can provide different insights into extreme values, such as dependency between extreme values, decaying rates, and return levels. Extreme spatial models are widely used in environmental statistics owing to the aforementioned advantages, while they cannot be directly applied to aircraft production systems owing to the different characteristics of the domains. 
Furthermore, because of the high sensitivity of the shape parameter in extreme models as well as various control inputs in aircraft production  \citep{lee2023partitioned}, it is worth exploring how to use extreme value modeling for aircraft production systems.  

We propose a multi-output extreme spatial model that enables extreme modeling of dynamic systems and the multi-output response thereof. The proposed model consists of two coupled models: (i) marginal parameter models for marginal distributions on the input space; and (ii) a max-stable process to model dependency between output variables. We define a proper metric space with geometric locations of critical points, where key characteristics of products are measured. The metric space enables the use of not only the max-stable process but also a graph structure of critical points. The undirected graph facilitates incorporating background knowledge and improving computational efficiency and estimation performance by utilizing truncated composite likelihoods. Our contributions in this paper can be summarized as follows. 
\begin{enumerate}
    \item A multi-output extreme spatial model for controllable production systems is introduced. The model captures patterns of multivariate extreme values conditional on control variables in a continuous design space.
    \item Graph-assisted composite likelihood estimation is used for high-dimensional inference. The graph structure facilitates the incorporation of background knowledge and geometric locations of critical points.
    \item End-to-end cutting-edge extreme modeling of composite fuselage assembly is demonstrated. The modeling process and application extensions for the usage in general production systems are provided. 
\end{enumerate}
Our case study on the composite fuselage shape control process in Section {\ref{se:case_study}} demonstrates that the multi-output extreme spatial model yields more promising  results for extreme cases, whereas the quantile regression models are less accurate than our model. Because accurate predictions of extreme values can improve the controllability over the production system or avoid significant losses due to system failures, the proposed model may benefit efficient production management.

The remainder of this paper is organized as follows. In Section \ref{se:mesm_literature}, we review related literature in the engineering domain that utilizes EVT and state-of-the-art extreme spatial models in other domains. In Section \ref{se:mesm_method}, fundamental EVT and extreme spatial models are provided, and extreme values in production systems are formalized. Thereafter, we introduce a multi-output extreme spatial model, including marginal parameter modeling, extremal dependence investigation, graph-assisted composite likelihood estimation, and computational algorithms. Section \ref{se:case_study} illustrates the application of the proposed method to extreme residual stress modeling in composite aircraft production. At last, the summary of this paper and further applications of the method are discussed in Section \ref{se:conclusion}.

\section{Literature Review}\label{se:mesm_literature}
In this section, we review recent literature on extreme value models in the engineering, data, and management sciences. Extreme values in engineering systems are usually associated with undesirable abnormal events, so they are mainly considered for system monitoring, failure prediction, and outlier detection. \cite{cho2019hierarchical} determined thresholds for anomaly detection with extreme value distributions. \cite{yu2021deep} used the extreme value model for the detection of outliers in open-set fault diagnosis and applied their method to the diagnosis of the rotating motor. Other studies have utilized extreme models for the predictive modeling of complex engineering systems. \cite{gu2020prediction} predicted the maximum fatigue indicator parameter in metal alloys, in which micro-scale defects are rare. In additive manufacturing, \cite{boyce2017extreme} used an extreme model to predict rare failures, which may cause significant loss of quality. \cite{yu2018time} utilized the extreme model for the prediction of multiple failure modes in time-variant reliability problems to reduce integral errors. The proposed method is applied to the assessment of the exoskeleton under uncertainty. \cite{liu2015statistical} modeled the thickness of the pipeline for the inspection of possible leakage therein. 

Another area where extreme values occasionally appear is reliability analysis, such as time-to-failure (TTF) prediction. \cite{xia2019online} used the extreme value distribution in the regression model of TTF for multimodal manufacturing machines. \cite{lewis2022prediction} used the Weibull distribution for the within-sample prediction of TTF in heterogeneous failure modes. \cite{king2016planning} used the Weibull distribution for the fatigue time of polymer composites and proposed the optimal design for variance-minimizing estimation. \cite{he2018predicting} used the extreme value distribution for modeling of first-failure-time given manufactured time. 
For cybersecurity, \cite{zhan2015predicting} used the nonstationary generalized Pareto distribution with time-varying parameters for the long-term prediction of extreme cyber attacks. However, their problem did not involve spatial dependency, so their approach cannot be directly applied to multi-output systems. \cite{grigoriu2017estimates} proposed the surrogate-assisted spatial extreme response model. They used a surrogate model with bounded noise for stochastic PDE and used the noisy samples for extreme response analysis. However, this method only utilizes the extreme value distribution for point-wise characterization without considering spatial extreme dependency.

There are works on extreme spatial models in other applications. \cite{davison2012statistical} provided a concise illustration of common spatial extreme models widely used and compared them in a rainfall case study. One active research area in extreme spatial modeling is to improve the flexibility and scalability of the model. The flexibility is associated with the asymptotic dependence between extreme events. Asymptotic dependence between extreme events means that the joint probability of extreme outcomes across multiple variables does not become zero as the threshold approaches, meaning the variables remain dependent in the tail region. While asymptotic independence represents the joint occurrence of extreme events becomes negligible in the tail region. Recently, asymptotic independent models and spatial processes with unknown extremal dependence class have received lots of attention \citep{huser2019modeling, bacro2020hierarchical, bopp2021hierarchical}.
 The scalability comes from the fact that many extreme spatial models are subject to the prohibitive computational cost of the full likelihood estimation because of the intractable density computation. \cite{castruccio2016high} compared performances of full likelihoods and composite likelihoods with different orders in relatively low-dimensional ($\le20$) problems and provided practical suggestions. \cite{engelke2020graphical} proposed the graphical models for multivariate Pareto distributions to exploit sparsity induced from the graph structure for efficient inference. 
Although numerous works have been done in this field, extreme models have not been well explored for the complex aircraft production systems.

\section{Methodology}\label{se:mesm_method}
To illustrate the multi-output extreme spatial model, we briefly review EVT and extreme spatial models as preliminaries. A mathematical formulation of production systems and the proposed multi-output extreme spatial model will be discussed afterward. For notations, capital letters denote matrices and functions, either univariate or multivariate, and lower-case letters are scalars. The dimension will be specified, e.g., $X\in\mathbb{R}^J$, and square brackets indicate the set of indices, e.g., $[J]=\{1,\ldots, J\}$. Lower letter subscripts denote component indices in vectors and sequences, and parenthesized superscripts indicate indices of i.i.d. samples. Some main notations that are frequently used in this paper are summarized in Table. {\ref{tab:notations}.}

\begin{table}[!t]
\TABLE{Major Notations and Definitions \label{tab:notations}}
{\renewcommand{\arraystretch}{1.1}
\centering
\begin{tabular}{M{0.05\textwidth}m{0.41\textwidth}M{0.05\textwidth}m{0.41\textwidth}}
\hline
$X$        & variable of interest                                        & $N$             & the number of design points                  \\
$Y$        & maxima of $X$                                 & $T$             & the block size for block maxima              \\
$Z$        & unit Fr\'echet transformed $Y$ & $B$             & the number of extreme values ($=\lceil L/T\rceil)$         \\
$\mu$      & location parameter                            & $\kappa_j$      & threshold for $j$-th critical point          \\
$\sigma$   & scale parameter                               & $\Psi$          & GP in marignal parameter model \\
$\xi$      & shape parameter                               & $\chi$          & extremal dependence                          \\
$V$        & exponent measure                              & $u_j$           & coordination of $j$-th critical point        \\
$s_n$      & $n$-th control input                          & $\tau$          & parameter of max-stable process              \\
$\Phi$     & multi-output production system                & $\eta$          & extremal coefficient                         \\
$\epsilon$ & observation noise                             & $\mathcal{G}_H$ & graph with order $H$                         \\
$\kappa_j$ & threshold for $j$-th critical point           & $q_G$           & quantile for graph                           \\
$L$        & the number of replications                    & $R$             & return level                                 \\ \hline
\end{tabular}}{}
\end{table}

\subsection{Preliminary: Extreme Value Models}\label{ss:mesm_evt_review}
Consider a random variable $X$ with a distribution $F_X$, and let $\{X^{(1)},\ldots, X^{(n)}\}$ be i.i.d. samples from $X$. Extreme value theory concerns the extreme value (i.e., maximum or minimum) of the random variable. For clarity in exposition, we confine our interest to the maximum without the loss of generality hereafter. The limiting maximum value of $X$ is
$$
    \lim_{n\to\infty} \frac{\max\{X^{(1)},\ldots, X^{(n)}\} - b_n}{a_n} \xrightarrow{D} Y,
$$
for some $b_n \in \mathbb{R}$ and $a_n > 0$ for $n\in\mathbb{N}$. Given that $Y$ has a nondegenerate distribution, the distribution is a member of 
\begin{equation}\label{eq:gev_family}
    F_Y(y) = 
        \exp\left[-\left\{1 +\xi \left(\frac{y-\mu}{\sigma}\right) \right\}^{-1/\xi}\right],
\end{equation}
which is defined on $\{y\colon 1 + \xi(y-\mu)/\sigma > 0\}$, where $\xi, \mu\in\mathbb{R}$ and $\sigma > 0$. The distribution family of \eqref{eq:gev_family} has three parameters and is called the generalized extreme value (GEV) family. The GEV family subsumes three subfamilies with different tail decaying behaviors according to the shape parameter $\xi$: (i) Weibull ($\xi < 0$); (ii) Gumbel ($\xi=0$); and (iii) Fr\'echet ($\xi>0$). An important property of the GEV family is that they are max-stable, which is defined to have some constants $a_n > 0$ and $b_n\in \mathbb{R}$ such that
$$
    F^n_Y(a_ny + b_n) = 
        F^{}_Y(y),
$$
for every $n\in\mathbb{N}$. Any distribution in the GEV family can be transformed into a unit Fr\'echet with
$$
    Z = 
        \left\{1 + \xi\left(\frac{Y-\mu}{\sigma}\right)\right\}^{1/\xi}\sim\mathrm{GEV}(1, 1, 1).
$$

Generally, the marginal transformation to unit Fr\'echet is mathematically convenient when one is interested in studying multivariate extreme values. Multivariate extreme value distributions with unit Fr\'echet margins can be written as
\begin{equation}\label{eq:multivariate_cdf}
    F_Z(Z_1\le z_1, \ldots, Z_J\le z_J) = 
        \exp\left\{-V(z_1, \ldots, z_J)\right\},
\end{equation}
where $V$ is called the exponent measure, which also characterizes the dependency between components. Developing a multivariate extreme value model with a flexible dependency is one active research area. The exponent measure can be either parametric or nonparametric, and must satisfy
$$
    V(\infty,\ldots,z_j,\ldots,\infty) =  
        1/z_j, \qquad 
    V(tz_1,\ldots, tz_J) = 
        t^{-1}V(z_1,\ldots, z_J),
$$
for any $t>0$ \citep{resnick2013extreme}. One of difficulties in multivariate extreme modeling is that we need partial derivatives of \eqref{eq:multivariate_cdf} for the likelihood estimation, which is computationally challenging concerning the dimension $J$.

Extreme spatial models are random processes defined over spatial domains in which marginals are extreme value distributions. An extreme spatial model can be defined with a smoothly varying model of marginal GEV parameters. However, it cannot address concurrent events at multiple locations in a spatial domain, so several copula models are proposed to introduce dependency \citep{sang2010continuous}. Max-stable processes are valid extreme spatial models since they represent random processes satisfying the max-stable property, and the widely used models include the Brown-Resnick process \citep{brown1977extreme} and the extremal-$t$ process \citep{opitz2013extremal}. Consider a spatial domain $\mathcal{S} \in \mathbb{R}^D$, where $s$ is an arbitrary location in $\mathcal{S}$. A max-stable process defined over $\mathcal{S}$ has the spectral representation \citep{de1984spectral}, 
\begin{equation}\label{eq:max_stable_model}
    Z(s) = 
        \sup_{i\in\mathbb{N}}\{W^{(i)}(s)/r^{(i)}\},
\end{equation}
where $r^{(i)}$'s are samples from the unit rate Poisson process on $\mathbb{R}_+$, and $W^{(i)}(s)$'s are from a stationary random process, satisfying $\mathbb{E}\left[\max\{0, W(s)\}\right]=1$. The random process characterizes $Z(s)$ with the corresponding exponent measure.

\subsection{Multi-output Extreme Spatial Model}
In this section, we illustrate the multi-output extreme spatial model for the complex aircraft production systems. Consider a multi-output production system, $\Phi\colon\mathcal{S}\to\mathcal{X}$, where $\mathcal{S} \subset \mathbb{R}^D$, and $\mathcal{X}\subset \mathbb{R}^J$. The input $s$ can be incoming states, control variables, or both, and the output $X$ is the resulting state or key characteristics of the product measured at $J$ critical points. The process can be modeled as
\begin{equation}\label{eq:process}
    X(s) = 
        \Phi(s) + \varepsilon(s),
\end{equation}
where $s \in \mathcal{S}$, and $\varepsilon \in \mathbb{R}^Q$ is observation noise. Many production systems are modeled with stochastic spatial models because of their intrinsic uncertainty attributed to unmeasurable or uncontrollable production variability. Operating production systems with high-dimensional observations is controlling the input to satisfy 
\begin{equation}\label{eq:criterion}
    X_j(s) < \kappa_j, \quad \forall j\in[J],
\end{equation}
where $\kappa_j$ is a tolerance for acceptable product quality. Generally, $\kappa_j$ is set to be large enough in product design to sustain the production process as much as possible, while keeping the product quality. Therefore, a promising way to monitor \eqref{eq:criterion} is focusing on extreme values of $X_j(s)$ that may exceed the tolerance.

\begin{figure}
    \FIGURE{
    \includegraphics[width=\columnwidth]{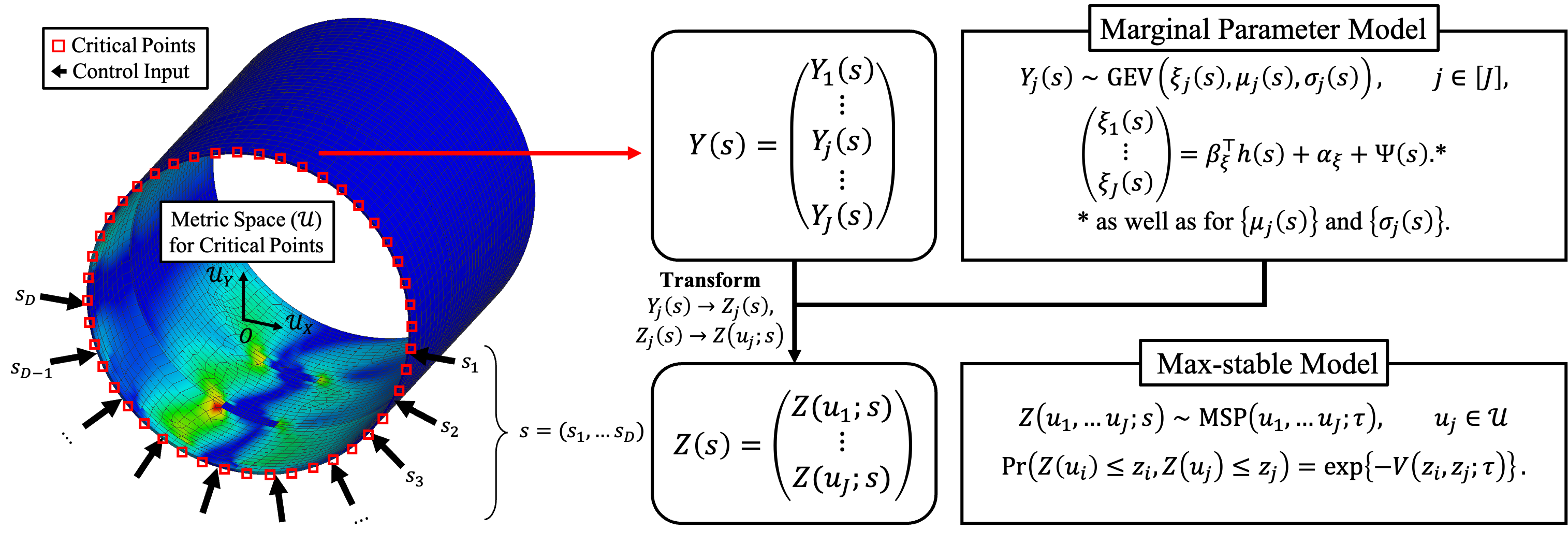}}
    {Overview of the Multi-output Extreme Spatial Model  for Aircraft Fuselage\label{fig:overview}}{}    
\end{figure}

The max-stable process in \eqref{eq:max_stable_model} cannot be readily used to model extreme values of \eqref{eq:process}, since $X$'s are conditionally independent over $\mathcal{S}$. Meanwhile, the marginal parameters of $X$ are spatially correlated over $\mathcal{S}$, so that $X_j(s)$ and $X_{j'}(s)$ may be correlated as concurrent events for $j,j'\in[J]$, where $j\ne j'$. Furthermore, many production systems involve a large number of critical points, which intensifies inferential computations in the max-stable process. The multi-output extreme spatial model addresses these drawbacks, and it is outlined in Fig. \ref{fig:overview} and Algorithm \ref{algo:moem}. The introduced model is in two steps. First, marginal parameter modeling estimates marginal parameters for control input with the extracted maxima from the original data, and constructs a spatial model over the input space to allow marginal modeling of extreme values. Second, multi-output modeling transforms the output into a unit Fr\'echet to employ a max-stable process to interconnect multi-output. A geometric space is defined to assign critical points in the space with a proper metric, and the distance between the critical points is used to construct a graph that accelerates the inference of the max-stable process. We will explain these steps and how to sample from the model in detail in the following sections.

\begin{algorithm}
\DontPrintSemicolon
\caption{Multi-output Extreme Spatial Modeling}\label{algo:moem}

\SetKwBlock{MarginalParameterModeling}{Step 1: Marginal Parameter Modeling (Section \ref{ss:marginal_parameter})}{}
\SetKwBlock{MultioutputModeling}{Step 2: Multi-output Modeling (Section \ref{ss:multi_output})}{}

\Input{
    $S,\ X,\ T,\ (\mathcal{U},\;d_{\mathcal{U}})$
    }
\MarginalParameterModeling{
    Extract maxima from $\{\boldsymbol{\mathrm{x}}(s_n)^{(l)}\colon l\in [L],n\in[N]\}$ with $T$ block size.\;
    \For{$n\in[N]$}{
        Get MLEs of the marginal parameters: $\boldsymbol{\mathrm{y}}(s_n) \sim \mathrm{GEV}(\Xi(s_n),\mathrm{M}(s_n),\Sigma(s_n))$.
        }
    Fit \eqref{eq:marginal_param_model} for each of $\{\Xi(s_n), \mathrm{M}(s_n), \Sigma(s_n)\colon n\in [N]\}$.\;
    }
\MultioutputModeling{
    \For{$(b,n)\in[B]\times[N]$}{
        Transform $\boldsymbol{\mathrm{y}}(s_n)^{(b)}$ into unit Fr\'echet margins, $\boldsymbol{\mathrm{z}}(s_n)^{(b)}$.
        }
    Define the metric space $(\mathcal{U},\, d_{\mathcal{U}})$ for $J$ critical points.\;
    Conduct extremal dependency investigation (e.g., with \eqref{eq:tail_coef1} and \eqref{eq:f_madogram}).\;
    Construct a graph $G_H$ with Algorithm \ref{algo:graph}.\;
    Fit the max-stable process $Z(\;\cdot\;;\tau)$ on $\{\boldsymbol{\mathrm{z}}(u_j)\colon j\in[J]\}$ maximizing \eqref{eq:tcl}.\;
    }
    
\Output{$\{\Xi(s), \mathrm{M}(s), \Sigma(s)\},\ Z(\;\cdot\;;\tau)$}
\end{algorithm}

\subsubsection{Marginal Parameter Modeling.}\label{ss:marginal_parameter}
Let $Y(s)$ be the component-wise maxima of $X(s)$, and denote $j$-th component of $Y(s)$ by $Y_j(s)$. We assume that $Y_j(s)$ is max-stable for all $s \in \mathcal{S}$ and $j\in[J]$ so that we can use the GEV family to model the marginal distributions of $\{Y(s)\colon s\in\mathcal{S}\}$. Given that we have $L$ replicated observations from $N$ design points $S=\{s_n\colon n\in[N]\}$, we may write observations as $\mathbf{X}(s_n) = \{\boldsymbol{\mathrm{x}}(s_n)^{(l)}\colon l\in [L]\}$ for each $n\in[N]$. Let $T$ be the block size for block maxima, which should be chosen based on backgrounds and inference stability. For example, the batch size, gross order, or delivery capacity can be considered in production systems. Otherwise, for inference stability, the block size can be chosen by looking at the return-level plot \citep{coles2001introduction} with the available data amount. Using block maxima with the block size $T$, extreme values of size $B=\lceil L/T\rceil$ can be extracted from observations as $\mathbf{Y}(s_n) = \{\mathbf{y}(s_n)^{(b)}\colon b\in[B]\}$. It leads to the estimation of GEV parameters using the maximum likelihood for each pair of $(n,j)\in [N]\times [J]$.

Let us denote the vector of each GEV parameter as $\Xi(s_n) = [\xi_1(s_n), \ldots, \xi_j(s_n)]^\top$, and $\mathrm{M}(s_n), \Sigma(s_n)$, respectively. We assume that $\xi_j(s_n)$, $\mu_j(s_n)$, and $\sigma_j(s_n)$ are independent for $j\in[J]$. We interpolate the marginal parameters of $\{\mathbf{Y}(s_n)\colon n\in[N]\}$ with
\begin{equation}\label{eq:marginal_param_model}
    \Xi(s_n; \beta_\xi, \alpha_\xi, \theta_\xi) = 
        \beta^\top_\xi f(s_n) + \alpha_\xi + \Psi(s_n; \theta_\xi),
\end{equation}
where $f:\mathcal{S}\to \mathbb{R}^I$ is a deterministic function of $s$ (e.g., polynomial), $\beta_\xi\in\mathbb{R}^{I\times J},\,\alpha_\xi\in\mathbb{R}^J$ are parameters for linear regression, and $\Psi$ is a $J$-dimensional zero-mean Gaussian process with hyperparameter $\theta_\xi$. In practice, the logarithm can be taken on $\Sigma(s_n)$ to ensure positiveness. Although components of $\Xi$ can be modeled independently via \eqref{eq:marginal_param_model}, using a multi-output GP \citep{bonilla2007multi} can be taken into account since parameters at geometrically close critical points may be correlated. However, the computational complexity of the multi-output GP is $\mathcal{O}(J^3N^3)$, so it can be prohibitive when $J$ is large. Therefore, the output dimension and the degree of parametric correlation should be considered in determining \eqref{eq:marginal_param_model}.

It is possible to utilize \eqref{eq:marginal_param_model} as a latent variable model (LVM) \citep{davison2012statistical} in a hierarchical model as
$$
    Y_j(s_n) \sim
        \mathrm{GEV}(\xi_j(s_n), \mu_j(s_n), \sigma_j(s_n)),\quad j\in[J],
$$
where proper priors are imposed on parameters. Although the LVM is straightforward, inference and sampling from the posterior distribution are computationally expensive when the control variable dimension is large, which is common in production systems. Moreover, adjusting a Markov chain Monte Carlo method for good convergence is intractable due to the high dimension of the parameter space. On the other hand, interpolating MLE parameters for $S$ is not hindered by the dimension as in the Bayesian framework. To sum up, the LVM is preferable when the control variable dimension is small; otherwise, interpolating the MLE parameters would compensate for the computational burden.

\subsubsection{Multi-output Modeling.}\label{ss:multi_output}
Although the marginal parameter model is sufficient to model the marginal extreme distribution for any $s\in \mathcal{S}$, it offers less information on the covariance between $J$ components in $Y(s)$. To model the dependency, we utilize the max-stable process in \eqref{eq:max_stable_model} with the following steps.
\begin{enumerate}
    \item Transform $Y(s)$ into vectors with unit Fr\'echet marginal distributions.
    \item Define a spatial domain for critical points.
    \item Investigate extremal dependency to select a proper max-stable process and to simplify the spatial domain.
    \item Estimate the parameter of the max-stable process by maximizing graph-assisted composite likelihood.
\end{enumerate}
The first step can be done component-wise since we have marginal distributions, and we denote the transformed output following the max-stable process parameterized with $\tau$ by $Z(s)\sim \mathrm{MSP}(s;\;\tau)$. The remaining steps are explained below.

\paragraph{Spatial Domain for Critical Points}
To model the intercorrelated $\{Z_j\colon j\in[J]\}$ with the max-stable process, the critical points should be represented in a spatial domain. Generally, measurement points in a production system are located on a rigid body, so it is reasonable to consider the geometric adjacency of critical points by defining a proper metric space, denoted by $(\mathcal{U},\, d_{\mathcal{U}})$. For a 2D rigid sheet, Euclidean distance can be measured on the surface of the sheet. Another way for complex component is to assign the $J$ critical points to $\{u_j\colon j\in[J]\}$ in a Riemannian manifold with the geodesic distance. Consequently, engaging the metric space associated with critical points transforms the max-stable process to a bilinear function on $\mathcal{S}\times \mathcal{U}$ as
$$
    \begin{bMatrix}{c}Z_1(s)\\\vdots\\Z_J(s)\end{bMatrix} \coloneqq
        \begin{bMatrix}{c}Z(u_1, s)\\\vdots\\Z(u_J, s)\end{bMatrix}\sim \mathrm{MSP}(u_1,\ldots, u_J, s;\; \tau),
$$
where the exponent measure is
\begin{equation}\label{eq:msp_measure}
    \pr{Z(u_1, s)\le z_1, \ldots, Z(u_J, s)\le z_J} = 
        \exp{\left\{-V(z_1, \ldots, z_J, s;\;\tau)\right\}}.
\end{equation}

\paragraph{Extremal Dependence Investigation.}
An important step in extreme spatial modeling is investigating extremal dependence to select a proper model. The extremal dependence in a bivariate case can be measured by
\begin{align}
    \chi &= \lim_{t\to 1}\pr{F_{Z_i}(z_i) > t|F_{Z_j}(z_j) > t}\label{eq:tail_coef1}\\
         &= 2 - V(1, 1),\label{eq:tail_coef2}
\end{align}
where $Z_i$ and $Z_j$ have unit Fr\'echet margins, and \eqref{eq:tail_coef1} can be empirically attained in practice. The coefficient $\chi$ measures asymptotic dependency between two extreme variables: $\chi=0$ indicates two variables are asymptotically independent, and $\chi=1$ means they are totally dependent. The estimated $\chi$ can be used to choose a proper extreme spatial model, and the model can be conversely validated with {\eqref{eq:tail_coef2}} if the exponent measure can be readily obtained. It is noteworthy that max-stable processes are asymptotically dependent, so one can employ corresponding models \citep{bopp2021hierarchical, huser2022advances} if data is presumed to be asymptotically independent.

Equation \eqref{eq:msp_measure} can be simplified by assuming that dependency is an intrinsic characteristic of the system, so the extremal dependence of $Z$ is independent of $s$ as
\begin{equation}\label{eq:cov_assumption}
    V(z_1,\ldots,z_J, s;\;\tau) \equiv
        V(z_1,\ldots,z_J, s';\;\tau),
\end{equation}
for any $s, s'\in\mathcal{S}$. To verify the assumption, we must show that $\chi$ is irrelevant to distance in $\mathcal{S}$. $F$-madogram \citep{cooley2006variograms} is useful to observe $\chi$ as a spatial dependence, which is 
\begin{equation}\label{eq:f_madogram}
    \eta(h) = 
        \ex{|F_{Z}(Z_j(s+h)) - F_{Z}(Z_{j'}(s))|} / 2,
\end{equation}
for $h>0$. Although \eqref{eq:cov_assumption} is not necessary in our model by defining the max-stable process on $\mathcal{S}\times \mathcal{U}$, it not only simplifies extremal dependence modeling but also alleviates data scarcity by transforming observations at different input locations in $\mathcal{S}$ as additional replications. It also allows the additional replications to be used for cross-validation in the parameter estimation by partitioning $S$.

\paragraph{Graph-assisted Composite Likelihood.}
For large $J$ (e.g., $J>20$), computation of the full likelihood is prohibitive since the joint density function of \eqref{eq:multivariate_cdf} involves all partial derivatives, of which the number of terms is the Bell number of $J$. Moreover, some max-stable processes even have no closed form. Therefore, a composite likelihood with a low-order (e.g., bivariate or trivariate) density is standard in the inference of max-stable processes. Consider the set of transformed observations $\mathbf{z}^{(b)} = \{z_j^{(b)}\colon j\in[J],\, \}$ for $b\in[NB]$. For the order $H\in\{2,\ldots, J\}$, $\mathcal{G}_H$ denotes the collection of all subvector of $\{\mathbf{z}^{(b)}\colon b\in[NB]\}$ with the size $H$, and $\boldsymbol{\mathrm{z}}_{[H]}$ is a member of $\mathcal{G}_H$. Let the parameter to be estimated in the max-stable process be $\tau$ and $f(\:\cdot\: ; \tau)$ is the $H$-th order density from \eqref{eq:multivariate_cdf} characterized by the max-stable process. Then, the conventional composite likelihood is 
\begin{equation}\label{eq:cl}
    \mathrm{CL}_H(\tau|\boldsymbol{Z}) 
        = \prod_{b=1}^B \prod_{\mathbf{z}_{[H]}\in \mathcal{G}_H} f(\mathbf{z}_{[H]}^{(b)};\tau),
\end{equation}
where the logarithm can be taken on the RHS as the composite log-likelihood.

However, even though a low-order composite likelihood is used, the computation can be demanding due to the size of $\mathcal{G}_H$, which is $C^J_H = \frac{J!}{H!(J-H)!}$. Truncated composite likelihoods use a nonempty subset $G_H \subset \mathcal{G}_H$ in \eqref{eq:cl} and can improve both the computation and the inference performance \citep{castruccio2016high}. A truncated composite likelihood can be written as
\begin{equation}\label{eq:tcl}
    \mathrm{TCL}_H(\tau|\boldsymbol{Z}, G_H) 
        = \prod_{b=1}^B \prod_{\mathbf{z}_{[H]}\in G_H} f(\mathbf{z}_{[H]}^{(b)};\tau),
\end{equation}
and can be regarded as an exponentially weighted composite likelihood in which weights are binary. The performance of the truncated likelihood estimation can be improved when $G_H$ consists of correlated components. Since key characteristics from geometrically neighboring critical points are likely to be correlated, the adjacency between points in each group can be a reasonable measure. In this paper, we refer to the average distance of the subvector, which is defined as
\begin{equation}\label{eq:distance}
    \delta(g) = 
        \frac{\sum_{u,u'\in g} d_{\mathcal{U}}(u, u')}{\mathrm{C}_2^H},
\end{equation}
where $g\in \mathcal{G}_H$ and $u\ne u'$. Ordering $\mathcal{G}_H$ with respect to \eqref{eq:distance} and taking a $q_G$-th quantile (or specifying an exact threshold based on background knowledge) will generate $G_H$ in which components are closely located. Algorithm \ref{algo:graph} is a pseudo-algorithm for generating $G_H$ for given $\delta$ and $q_G$. The subcollection $G_H$ can be regarded as an undirected graph in which cliques have the size $H$, so we refer to \eqref{eq:tcl} with $G_H$ generated in this way as a graph-assisted composite likelihood. 
\begin{algorithm}
\DontPrintSemicolon
\caption{Generating Graph}\label{algo:graph}
\Input{$\mathcal{G}_H,\ \delta,\ q_G$}
$\Delta = \emptyset$\;
\For{$g\in \mathcal{G}_H$}{
    $\delta(g) = \sum_{u,u'\in g} d_{\mathcal{U}}(u, u')/\mathrm{C}_2^H\ \mathrm{for}\ u\ne u'$\;
    $\Delta = \Delta \cup \{\delta(g)\}$    
}
Estimate $q_{G}$-th quantile of $\Delta$ as $\Delta_{q_{G}}$\;
$G_H = \{g\colon g\in \mathcal{G}_H, \delta(g)\le \Delta_{q_{G}}\}$\;
\Output{$G_H$}
\end{algorithm}

\subsubsection{Sampling}\label{ss:sampling}
Once the model is fitted by maximizing the graph-assisted composite likelihood, we can sample extreme values from the proposed model for unobserved $s\in\mathcal{S}$ with the following two steps.
\begin{enumerate}
    \item Sample $z(s)$'s from $Z(u_j,\;\tau)$ for $j \in [J]$.
    \item Inverse transform $z(s)$'s to $y(s)$'s with $\{\Xi(s),\;\mathrm{M}(s),\;\Sigma(s)\}$.
\end{enumerate}

Sampling $z$'s from the max-stable process in the first step is not as simple as other statistical models since \eqref{eq:max_stable_model} requires infinite samples from the spectral representation. Although it is possible to take the maximum of large samples, it is beneficial to use the exact simulation method proposed by \cite{dombry2016exact}. It facilitates more efficient sampling using conditional spectral random processes, to which we resort in the case study.

\section{Simulation Study}\label{se:simulation_study}
In this section, we empirically study the parameter recovery of the graph-assisted composite likelihood method with a simulation study. To generate extreme values, the Brown-Resnick model \mbox{\citep{brown1977extreme}} is defined over a unit square, $\mathcal{U} = [0,\, 1]^2$, and the true parameter of the radial basis kernel is set to $\tau = (0.5,\, 0.02)$. $J=20$ critical points are uniformly drawn over $\mathcal{U}$, and $B=20$ extreme value samples are generated. As an analogy, the simulation study can be regarded as measuring residual stress over a metal sheet to detect extremely large stress as an anomaly. The critical points are where residual stress is measured, and the measurement is repeated 20 times, considering measurement error.

The graph-assisted composite likelihood is used to estimate the parameters of the multi-output extreme spatial model, and we consider different $q_G\in\{0.2,\,0.4,\,0.6,\,0.8,\,1.0\}$, to observe its effects. Note that the samples at each critical point are independently drawn from a unit Fr\'echet, so we do not employ the marginal parameter model in this simulation. 

The simulation is implemented 20 times to mitigate the effect of randomly selected critical point locations. For the evaluation, we consider $L_1$-distance between the maximum composite likelihood estimation and the ground truth,
\begin{equation*}
    \text{Score}=
        \|\tau - \tilde{\tau}\|_1,
\end{equation*}
where $\tilde{\tau}$ is the estimated parameter, and the total time taken in parameter estimation.

\begin{figure}
    \FIGURE{
    \includegraphics[width=0.5\textwidth]{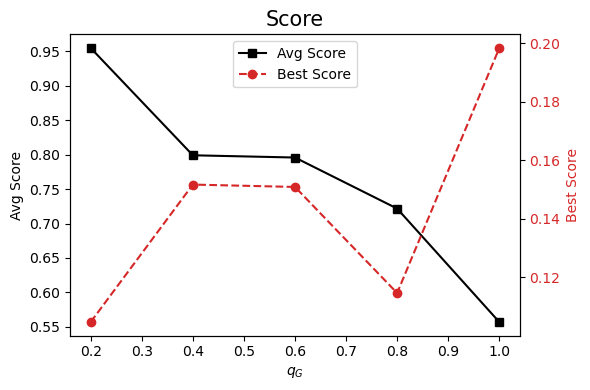}\hfill
    \includegraphics[width=0.5\textwidth]{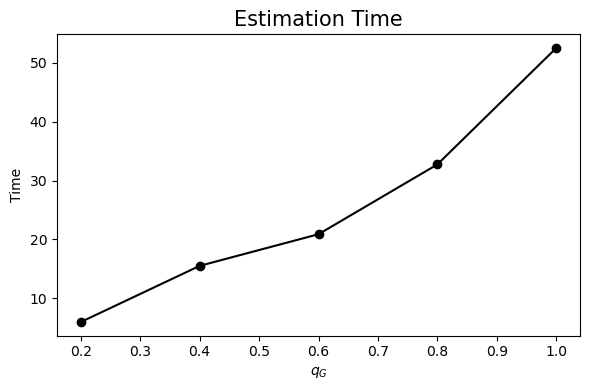}}
    {Simulation Results with Different $q_G$\label{fig:simulation}}
    {}
\end{figure}

Fig. \ref{fig:simulation} shows the simulation results. The left figure shows the average and best scores among 20 replications with different critical points. The best scores are less than 0.2, which implies that estimation with maximizing graph-assisted composite likelihood performs well. However, increasing $q_G$, so that the composite likelihood involves more pairs, does not necessarily induce better estimation. Meanwhile, the average score shows that a larger $q_G$ shows better stability in parameter estimation. The estimation time in the right figure demonstrates that higher $q_G$  values correspond to increased computing time by involving more terms in calculating  {\eqref{eq:tcl}}. To summarize, we can see that the graph-assisted composite likelihood, properly truncated with $q_G$, may improve information relevancy and reduce computation time by discarding distant locations. It is subject to sampling locations, and involving more terms in the composite likelihood becomes robust to the variability.

\section{Case Study: Composite Aircraft Production}\label{se:case_study}
This section applies the proposed method to model maximum residual stress in composite aircraft production, specifically, fuselage assembly. Briefly, due to inevitable dimensional deviations, the assembly process involves fuselage shape control using hydraulic actuators on fuselage sections installed on fixtures. Once the fuselage sections are reshaped into desired dimensions, they are assembled using rivet joints; this sequence of procedures causes residual stress in the structure as shown in Fig. \ref{fig:overview}. More details on the fuselage assembly process can be found in \cite{wen2019virtual}. The characteristics of the assembly process can be summarized as follows, which are also common in modern manufacturing systems.
\begin{itemize}
    \item The process is subject to intrinsic uncertainty due to actuator load noise, fixture locations, material inhomogeneity, etc. These factors complicate modeling solely based on engineering knowledge.
    \item Deformation and residual stress in composite parts are complicated because composite materials are composed of different materials and multiple layers with different orientations. Generally, their behaviors in response to external effects are nonlinear and anisotropic.
    \item The process yields inevitable residual stress in composite parts after releasing the actuators from the fuselage. Extremely large stress induced by improper actuator loads may lead to structural failures (e.g., fracture, brittle).
    \item The quality of the assembled fuselage is related to safety issues, so ultra-high precision control is required, which entails a rigorous inspection with a large number of repetitive measurements and sensors. It compels high-dimensional data analytics.
\end{itemize}
The traditional shape control process requires proficient experts and computationally prohibitive simulations, so surrogate modeling, Bayesian optimization, and active learning methods are proposed to address the aforementioned challenges \citep{yue2018surrogate, albahar2022physics}. For structural failures, \cite{lee2022physcal} considered residual stress to prevent failures in surrogate modeling, while they referred to deterministic simulation with the GP, which is unrealistic in practice due to intrinsic uncertainty. 

\subsection{Experiment Setting}
In this case study, we used the surrogate model from \cite{lee2022neural} as the ground truth to simulate composite fuselage assembly with $D = 20$ actuators (10 actuators for each section) and residual stress measured at $J = 128$ equispaced critical points around the joint rim. The actuators are subject to intrinsic Gaussian noise where the mean is zero, and the variance is 5\% of the force magnitude. For training samples, $N = 30$ design points are selected with the maximin Latin Hypercube design \citep{santner2018design} over $\mathcal{S} = [-200, 200]^{20}$ (lbf), and the simulation is replicated $L=500$ times for each design point. Consequently, we have the design matrix $S$ with the size $30\times 20$ and observations with $30\times 500 \times 128$. For testing, we select 20 design points in the same way as the training data. To extract maxima, different block sizes are considered to see the effect based on the number of samples and the reported yearly orders of one major composite aircraft, Boeing 787, which roughly varies between 25 and 350. For the metric space for critical points, we defined $\mathcal{U}\subset \mathbb{R}^2$ where the origin is the center of the nominal joint planar fuselage (see Fig. \ref{fig:overview}). For $d_{\mathcal{U}}$, the shortest path along the surface between two points is considered.
\begin{figure}
    \FIGURE{
    \includegraphics[width=\textwidth]{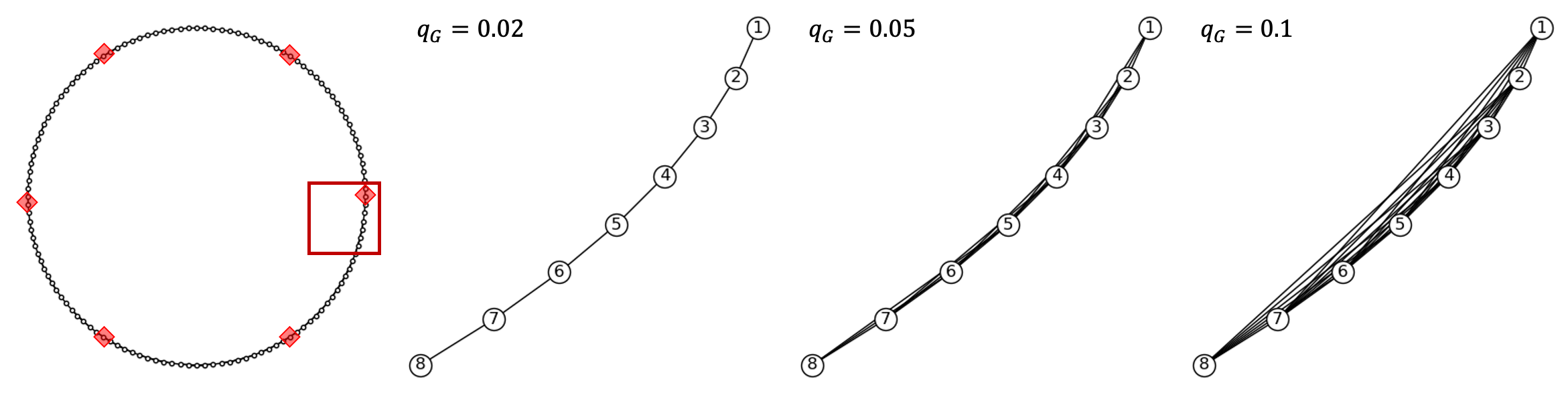}}
    {Graphs of Critical Points with Different Quantiles\label{fig:graph}} 
    {The critical points highlighted with diamonds are the representative critical points that are used in evaluating extremal dependence (i.e., 6 out of 128). The three right figures show subgraphs with 8 critical points in the red box. As $q_G$ increases, more cliques (connected nodes) are considered in the composite likelihood. $q_G = 0.02$ is used in the case study.}
\end{figure}

\begin{figure}
    \FIGURE{
    \includegraphics[width=0.32\textwidth]{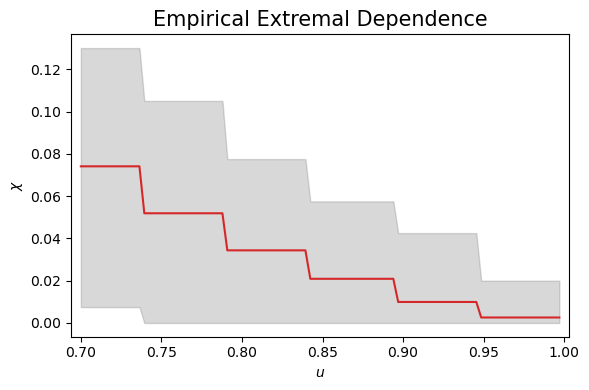}
    \includegraphics[width=0.32\textwidth]{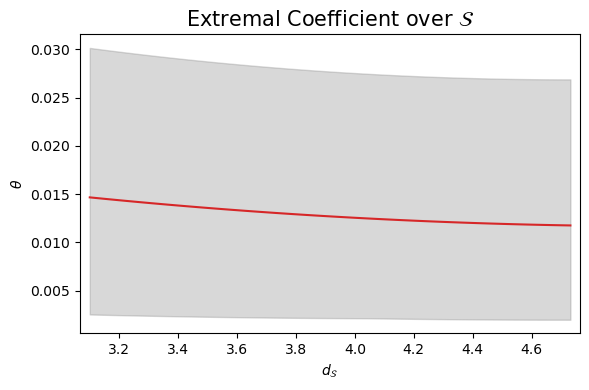}
    \includegraphics[width=0.32\textwidth]{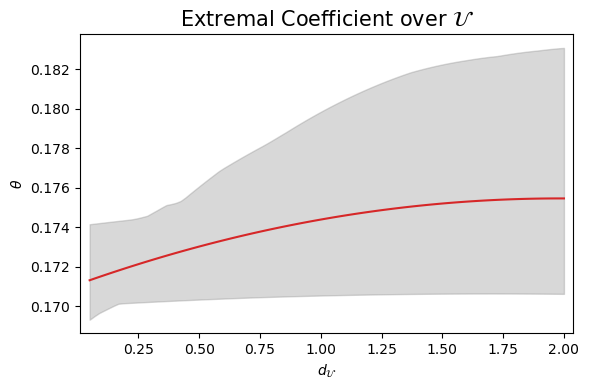}}
    {Extremal Dependence Investigation on Residual Stress with 95\% Confidence Intervals from Bootstrapping.\label{fig:dependence}}
    {}
\end{figure}

To investigate extremal dependence in maximum residual stress, we refer to empirical estimates of \eqref{eq:tail_coef1} between points as shown on the left of Fig. \ref{fig:dependence}. To reduce the number of pairs for evaluating \eqref{eq:tail_coef1}, six representative critical points, exhibiting large variance, are chosen as shown in Fig. \ref{fig:graph}. Even though the considered critical points include uttermost points, the empirical value of $\chi$ varies up to 0.02 with an average of 0.0025 as shown in the left figure in Fig. \ref{fig:dependence}. Therefore, extreme values between adjacent critical points may be dependent. The two right figures in Fig. \ref{fig:dependence} show empirical $F$-madograms with respect to Euclidean distance in the actuator design space and the metric space of critical points, respectively. The $F$-madogram with respect to actuator force shows that the extremal dependence barely changes in the design space, which validates \eqref{eq:cov_assumption} on $\mathcal{S}$. Meanwhile, the $F$-madogram for the critical point distance presents a usual dependence behavior, which increases with the distance. In conclusion, the exponent measure in \eqref{eq:cov_assumption} is valid in this case, and the defined metric space $\mathcal{U}$ is proper for the extreme spatial model.

To compare the proposed method with canonical predictive models in composite fuselage assembly, we consider quantile stochastic kriging (QSK) \citep{davison2012statistical, Plumlee2024}, stochastic kriging (SK) \citep{ankenman2010stochastic} and quantile linear regression (QLR) with different quantiles: $q\in\{0.9,\, 0.95,\, 0.99\}$. The QSK is trained with data over $q$-th quantile, while the SK and the QLR are trained with all observations (i.e., including non-extreme data). Also, to see the effect of the block size in our model, we considered different sizes: $T\in\{10,\, 20,\, 25,\, 50\}$. For the proposed model, we used \eqref{eq:marginal_param_model} as the marginal parameter model, and the Brown-Resnick process as the max-stable process for multi-output modeling, respectively. For SK, QSK, and the GP in the marginal parameter model, the radial basis function (RBF) kernel with the scale, length, and nugget parameters is considered, and the parameters are estimated with maximum likelihood using the quasi-Newton method. Meanwhile, the Brown-Resnick process used the RBF kernel without scale and nugget parameters (i.e., $\tau$ is a 2-D length parameter). The parameters of the QLR are estimated using the linear optimizer in \texttt{scipy} \citep{scipy2020nature} on the pinball loss associated with the quantile. To fit the max-stable process, the bivariate density ($H=2$) is used for the truncated likelihood with the graph with quantile $q_{G}=0.02$, which consists of all adjacent critical point pairs as shown on the left of Fig. \ref{fig:graph}. The number of terms in the truncated composite likelihood is 162 (i.e., 2\% of $=C^{128}_2)$), while the non-truncated likelihood consists of 8,128 terms. It is noteworthy that it is possible to introduce further prior knowledge to the graph, such as fixture or actuator locations that may influence the adjacent dependency. However, we did not consider such information in this case study.

For model evaluation, we consider two metrics because the QLR's distributional prediction is not as straightforward as other methods, and the techniques are beyond the scope of this paper. As the major metric, we use the empirical Wasserstein distance (WD) between the testing dataset and samples from the marginal distributions of each model to comprehensively compare the distributional distance. The empirical distance between two random variables is defined as
$$
    \mathrm{WD}(Y_1, Y_2) = 
        \sum_{n=1}^N\frac{|Y_1^{(n)} - Y_2^{(n)}|}{N},
$$
where $Y_1^{(n)}$ and $Y_2^{(n)}$ are ordered samples from $Y_1$ and $Y_2$, respectively. Since the metric is subject to sampling variability from the marginal distribution, we resample from the marginal distribution as many times as the number of testing datasets, and use bootstrapping for the testing dataset to refer to the average distance. In its own right, the WD metric is appropriate for evaluating the discrepancy between two distributions, which is frequently used in distributionally robust optimization. The second metric is the percentage mean distance (PMD), which refers to the point estimation of the mean as
$$
    \mathrm{PMD}(Y_1, Y_2) = 
        \frac{|\bar{Y_1} - \bar{Y_2}|}{|\bar{Y_1}|},
$$
where $\bar{Y}$ is the arithmetic mean of $Y$. Although it is not as comprehensive as the WD, it facilitates comparing the QLR to other methods in terms of extremal trends in residual stress.

\subsection{Results}
\subsubsection{Predictive Performance.} 
Table \ref{tab:result} summarizes the results of the considered methods, where the MESM stands for the multi-output extreme spatial model, the proposed method. The scores for all critical points are averaged, and the testing dataset and the standard deviations are calculated using bootstrapping. The result shows that the MESM outperforms the other models in WD regardless of the block size. The SK resulted in insufficient predictive accuracy in both metrics, since it does not place as much weight on extreme values as the other methods. With the PMD, the QSK performed the best, while the QSK is inferior to the MESM in fully describing the extreme stress distributions. Moreover, we can observe that the QLR is even worse than the MSEM in the PMD. 

\begin{figure}
    \FIGURE{
    \includegraphics[width=0.5\textwidth]{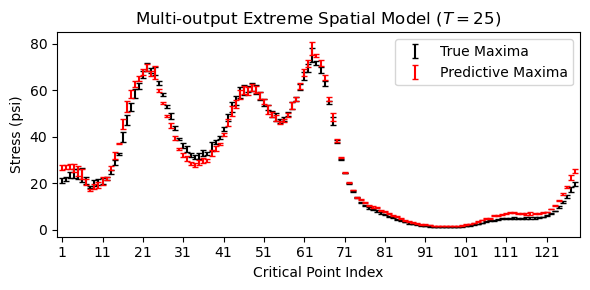}\hfill
    \includegraphics[width=0.5\textwidth]{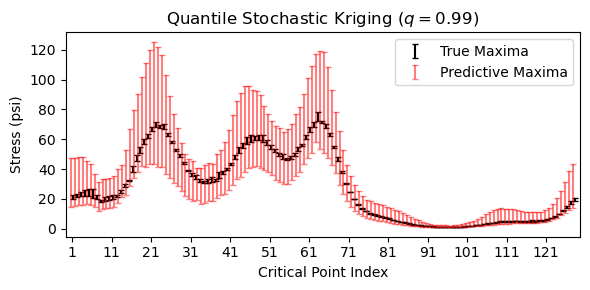}}
    {95\% Confidence Intervals of Marginal Distributions of Extremal Residual Stress in Fuselage Assembly\label{fig:marginal_prediction}}
    {}
\end{figure}

Fig. \ref{fig:marginal_prediction} shows the marginal predictive distributions of the MESM and the QSK for an instance of extremal residual stress from the testing dataset, obtained using the block maxima with $T=25$. It clearly shows that the MESM describes the behavior of the extremal residual stress distribution well, while the QSK has too large intervals, which is implausible. Also, it shows that residual stress tends to be high at the bottom side of the fuselage due to the fixture and direct contact with the actuators.

\begin{table}
\TABLE
{Averaged Distances on Testing Dataset and Training Times of Models\label{tab:result}}
{\renewcommand{\arraystretch}{1.3}
\resizebox{\columnwidth}{!}{%
\begin{tabular}{cccccccccccc}
\hline
Models &
  \multicolumn{4}{c}{MESM} &
  \multicolumn{3}{c}{QSK} &
  \multicolumn{3}{c}{QLR} &
  SK \\ \hline
Parameter &
  $T=10$ &
  $T=20$ &
  $T=25$ &
  $T=50$ &
  $q=0.90$ &
  $q=0.95$ &
  $q=0.99$ &
  $q=0.90$ &
  $q=0.95$ &
  $q=0.99$ &
  - \\ \hline
WD &
  \begin{tabular}[c]{@{}c@{}}3.555\\ (2.396)\end{tabular} &
  \begin{tabular}[c]{@{}c@{}}3.572\\ (2.376)\end{tabular} &
  \textbf{\begin{tabular}[c]{@{}c@{}}3.553\\ (2.399)\end{tabular}} &
  \begin{tabular}[c]{@{}c@{}}3.684\\ (2.562)\end{tabular} &
  \begin{tabular}[c]{@{}c@{}}5.965\\ (3.989)\end{tabular} &
  \begin{tabular}[c]{@{}c@{}}5.948\\ (3.956)\end{tabular} &
  \begin{tabular}[c]{@{}c@{}}5.911\\ (3.933)\end{tabular} &
  - &
  - &
  - &
  \begin{tabular}[c]{@{}c@{}}8.901\\ (10.187)\end{tabular} \\
PMD &
  \begin{tabular}[c]{@{}c@{}}0.150\\ (0.046)\end{tabular} &
  \begin{tabular}[c]{@{}c@{}}0.161\\ (0.096)\end{tabular} &
  \begin{tabular}[c]{@{}c@{}}0.152\\ (0.047)\end{tabular} &
  \begin{tabular}[c]{@{}c@{}}0.158\\ (0.069)\end{tabular} &
  \textbf{\begin{tabular}[c]{@{}c@{}}0.112\\ (0.024)\end{tabular}} &
  \begin{tabular}[c]{@{}c@{}}0.114\\ 0.026)\end{tabular} &
  \begin{tabular}[c]{@{}c@{}}0.121\\ (0.029)\end{tabular} &
  \begin{tabular}[c]{@{}c@{}}0.169\\ (0.057)\end{tabular} &
  \begin{tabular}[c]{@{}c@{}}0.182\\ (0.064)\end{tabular} &
  \begin{tabular}[c]{@{}c@{}}0.203\\ (0.070)\end{tabular} &
  \begin{tabular}[c]{@{}c@{}}0.121\\ (0.018)\end{tabular} \\
\begin{tabular}[c]{@{}c@{}}Training\\ Time (min.)\end{tabular} &
  17.105 &
  17.961 &
  18.518 &
  17.621 &
  0.509 &
  0.484 &
  0.494 &
  50.679 &
  50.193 &
  32.588 &
  0.402 \\\hline
\end{tabular}%
}
}
{\textit{Note. }Parenthesized numbers are standard deviations.}
\end{table}

\subsubsection{Training Time.} 
Most of the time in training the MESM was taken by fitting the max-stable process for multi-output modeling, which took about 15 minutes with the graph with $q_{G}=0.02$ and resulted in $\tau=(10.569,\, 11.126)$. Meanwhile, training the max-stable process with the full composite likelihood took about 180 minutes, resulting in $\tau=(11.911,\, 11.795)$. Behaviors of the sample paths from the two models were negligible as shown in Fig. \ref{fig:case_sample}. The QSK and the SK took less than a minute due to their lack of dependence modeling. The QLR also does not include the dependence model, but it took the most time among the compared models.

\subsection{Application Extensions}
\begin{figure}[ht]
    \FIGURE{
    \includegraphics[width=0.5\textwidth]{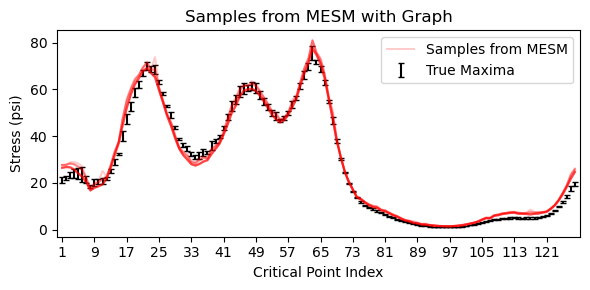}\hfill    
    \includegraphics[width=0.5\textwidth]{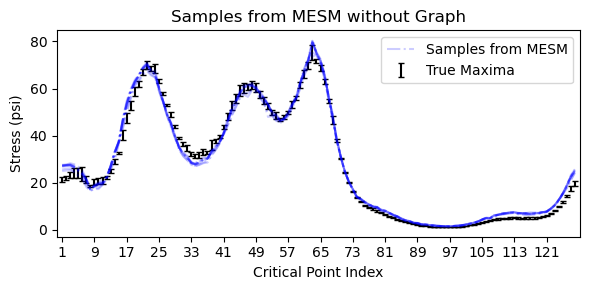}}
    {Simulations with the MESM\label{fig:case_sample}}
    {Ten samples are drawn from each MESM.}
\end{figure}

The extreme spatial model may provide useful insights for extreme values in aircraft production systems. The first thing that we can do is simulate extreme events in the assembly process. Samples from the extreme model represent the maximum stress trend in the process, and it may be used to efficiently monitor the assembly process by allocating sensors to a few vulnerable critical points. For example, an observed actuator input is expected to yield extremal residual stress in Fig. \ref{fig:case_sample}, which produces the largest stress at the critical point \# 62, to which the manufacturer should refer with more caution. 

\begin{figure}[ht]
    \FIGURE{
    \includegraphics[width=0.5\textwidth]{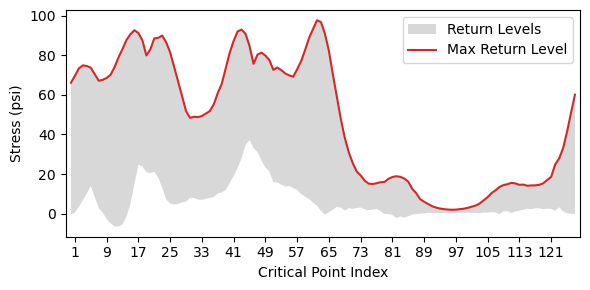}}
    {Return Levels ($R=100$) on the Control Design Space\label{fig:return_level}}{}
\end{figure}

A further promising application is to use the $R$-return level of a GEV distributed value $Y$ in the design and control of the production system, robust to extreme risks. The $R$-return level for $s\in \mathcal{S}$ at $j$-th critical point is defined as 
\begin{equation}\label{eq:return_level}
    \mathrm{RL}_j(s;R) =
        \mu_j + \frac{\sigma_j}{\xi_j}\{(-\log p)^{-\xi_j} - 1\}, \quad p\in(0, 1),
\end{equation}
where $p=1 - 1/R$ and $R \in \mathbb{N}$. Equation \eqref{eq:return_level} can be interpreted as the value $Y$ exceeding $y_p$ once on average every $R$ cycles of the block size (e.g., lot size, batch size, yearly deliveries). Fig. \ref{fig:return_level} illustrates $R=100$ return levels for all $s\in\mathcal{S}$, and the red line is the maximum. The maximum is at most 100 psi, which is far less than the yield stress of the composite material, so that the assembly process may utilize higher actuator loads for efficiency. If the system is designed and controlled based on return levels, the manufacturer may also save the inspection cost by concentrating the monitoring assets at the most vulnerable critical points or locations.

\section{Conclusion}\label{se:conclusion}
Extreme events in production systems are rare but significant. While statistical modeling is common in production management, extreme spatial models have been underutilized, even though many undesirable outcomes stem from tail distributions. We introduced a multi-output extreme spatial model for production systems that is capable of modeling correlated key characteristics and extremal dependence. The proposed model establishes the marginal parameter model over the control input space to estimate marginal distributions, and the max-stable process is used to link key characteristics based on the geometric locations of critical points and associated with the asymptotic dependence property. We also developed graph-assisted composite likelihood estimation to incorporate background knowledge and improve the estimation efficiency. 

The case study result shows that the multi-output extreme spatial model can be a promising option to capture extreme residual stress in composite fuselage assembly with high accuracy, when the conventional approaches, such as quantile regressions, are not sufficient. Furthermore, simulation with the proposed model is demonstrated, and the application extension for aircraft assembly operation is discussed. Apart from the above, extreme risk-robust control can also be a promising application with the extreme spatial model by embedding the model into constraints in optimization policy updates. Although the case study mainly considered composite aircraft production, the proposed approach also has the potential to be applied to other dynamic systems prone to extreme events.

\section*{Data and Code}
The data and the code for implementation will be available upon request. 

\ACKNOWLEDGMENT{%

The work was partially supported by the the National Natural Science Foundation of China (No. 92467302), Beijing Natural Science Foundation (No. L241039) and the Opening Project Fund of Materials Service Safety Assessment Facilities. 

}



%
%
%


\bibliographystyle{informs2014} 
\bibliography{bib_lee2} 


\end{document}